\documentclass[pre,floats,aps,amssymb,preprint,epsf,epsfig,rotate]{revtex4}
\usepackage{graphicx}
\begin{document}

\title{Experimental evidence on the development of scale invariance
in the internal structure of self-affine aggregates.}

\author{C. M. Horowitz, M. A. Pasquale, E. V.
Albano and A. J. Arvia}

\affiliation{Instituto de Investigaciones Fisicoqu\'{\i}micas Te\'{o}ricas
y Aplicadas, (INIFTA), CONICET, UNLP.
Sucursal 4, Casilla de Correo 16, (1900) La Plata. ARGENTINA. FAX : 0054-221-4254642.
E-mail : horowitz@inifta.unlp.edu.ar}



\begin{abstract}
It is shown that an alternative approach for the characterization of  
growing branched patterns consists of the statistical analysis of 
frozen structures, which cannot be modified by further growth, that 
arise due to competitive processes among neighbor growing
structures. Scaling relationships applied to these structures provide 
a method to evaluate relevant exponents and to characterize growing
systems into universality classes. The analysis is applied 
to quasi-two-dimensional electrochemically formed silver branched
patterns showing that the size distribution of frozen structures
exhibits scale invariance. The measured exponents, within the error bars, 
remind us those predicted by the Kardar-Parisi-Zhang equation.  
\end{abstract}

\pacs{PACS numbers: 68.35.Ct, 05.40.-a, 02.50.-r, 81.15.Aa }

\maketitle

\pagebreak

Interfaces grown under nonequilibrium conditions often show  scaling behaviors
and have attracted considerable interest\cite{2,3}. An intense
theoretical activity based on the study of continuum as well as discrete growth
models has provided scaling relationships between growth front
roughness (i.e., the mean square fluctuation of interface position) and
thickness, and upon changing the spatial scale
of observation. Structures that preserve a similar morphology
under an anisotropic magnification are termed self-affine.
They play a key role in
the rationalization of growth modes and the development of  
theories for growing aggregates. For a self-affine surface growing in
a confined geometry of typical side $L$, the interface roughness ($W$)
follows simple power-laws: $W \propto L^{\alpha}$ for $t \gg t_{c}$ 
and $W(t) \propto t^{\beta}$ for $t \ll t_{c}$, where $t_{c} \propto
L^z$  is the crossover time between these two regimes. The scaling
exponents  $\alpha$, $\beta$ and $z $  are called roughness,  growth and dynamic exponents,
 respectively,  and they are not 
independent, actually $z = \alpha/\beta$\cite{vic}. These exponents
allow us to identify the universality class of the system. In this context, 
experimental results from a large number of very different systems have been
reported \cite{2}.

In contrast to the huge theoretical and experimental effort devoted to
the understanding of the self-affine nature of interfaces, little
attention has been drawn to the study of the internal (bulk) structure
of growing systems. This approach is based on the fact 
that any growing system can effectively be rationalized on the basis
of a treeing process; i.e., any growing structure can be thought as 
the superposition of individual trees\cite{RV,Meakin0,zinc,Meakin,MM,Krug}.
 Those trees that spread out 
incorporating additional growing centers e.g., capturing particles, 
developing new branches, are said to be alive. In contrast, other
trees that may stop growing due to shadowing by
surrounding growing trees are termed dead trees.  The structure of dead
trees remains frozen because it cannot be modified by any further
growth. Considering that relationships among exponents relevant for
the distribution of dead trees and the standard scaling exponents have been
established theoretically and tested numerically \cite{RV,Meakin0,zinc,Meakin,MM,Krug,trees},
this approach may provide an alternative (experimental) method for
characterizing growing systems. Thus, this method
could be used in any experimental situation where the competitive 
dynamics of growing
trees leading to frozen structures can be observed.
A promising application of this method could
be  the electrochemical deposition of metals in thin cells. This rather simple
experimental technique  offers the possibility to observe a variety of
 self-affine patterns,  ranging from dense branching  to
diffusion-limited aggregation  
\cite{matsu1,matsu2,barkey,ec1}. Another application could be
to study polycrystalline thin film growth by scanning tunneling  microscopy
\cite{stm1,stm2} and atomic force microscopy \cite{afm}. In these cases the
direct evaluation of the number of crystallites as a function of the deposit
height would account for the number of frozen crystallites (i.e., the trees of
the growing aggregate) and the crystallite size distribution can then be evaluated.
Also, self-organized patterns formed upon bacterial colony growth,
which can be rationalized in terms of competing knotted-branched 
structures \cite{natu}, may provide an interesting application of
the proposed method.

Despite the fact that internal structure analysis has been developed
theoretically and tested numerically in various growing models,
 its application to real systems is very poor in the literature.  
The goal of this paper is to fill this gap by
characterizing quasi-two-dimensional electrodeposited silver patterns
produced under mass transfer kinetic control. Silver
electrodeposition was carried out at two different cathode-anode 
potentials, under the same kinetic control mechanism. 
Although long range interactions are present in a number of metal electrodeposition 
processes, in previous work it was shown that the standard exponent $\alpha$, $\beta$ and $z$ derived 
from growth patterns obtained under conditions similar to those reported here, 
asymptotically approach the values predicted by the Kardar-Parisi-Zhang (KPZ) 
universality class \cite{ec2,m1}. In this work we show that the internal structure statistical 
analysis of silver electrodeposits renders scaling relationships and relevant exponents that, 
within the error bars, also remind us those exponents predicted by the KPZ universality 
class.   

Silver electrodeposition was performed using a quasi-2D rectangular cell
with a vertical parallel plate electrode arrangement. The cell
consisted of two optical glass plates placed horizontally and separated by
0.025 cm. The cathode and anode were made of a silver foil, 99.9 \%
purity, 0.025 cm thick. The cathode width was 9 cm and the cathode to anode
distance was 4 cm. Further details of the experimental setup
 can be found in Ref.\cite{m1}. 
Aqueous solution 0.01 M silver sulfate / 0.5 M sodium sulfate /
0.01 M sulfuric acid, saturated with nitrogen, freshly prepared
utilizing  analytical grade
reagents and Milli-Q water, was employed. The electrodeposition was run 
under two different constant cathode-anode potentials, 
i.e., $\Delta E _ {c-a}  = - 0.80 $ V,
and  $\Delta E _ {c-a} = -1.20 $ V, employing a Radiometer
Voltamaster 32 potentiostat. Images of whole growth patterns were acquired
 at the end of each experiment using a
charge coupled device video camera, Hitachi 220, coupled to a computer
equipped with a frame grabber and image analyzer, Contron
Electronics KS. After binarization, the spatial resolution of the images 
was $31.25 \mu m$ per pixel. A program written in Fortran was used 
to distinguish internal trees separately and
to evaluate the distribution of dead trees.

\begin{figure}
\centerline{
\includegraphics[width=8cm,height=7cm,angle=0]{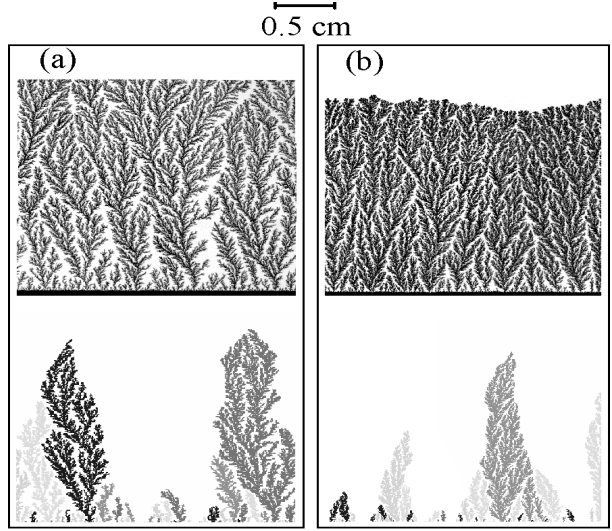}}
\caption{Silver pattern electrodeposits from 0.01M silver sulfate / 0.5 M
sodium  sulfate / 0.01 M sulfuric acid.  (a) at $\Delta E_{c-a} = - 0.80 $ V, electrodeposition time 110 min and (b)  at  $\Delta E_{c-a} = - 1.20 $ V,
electrodeposition time 90 min . The dead trees corresponding to both
patterns are  shown at the bottom of each figure}
\label{Figure1} 
\end{figure}

Silver electrodeposition is one of the fastest electrochemical reactions due to 
its high exchange current density value \cite{g1}.
  Despite the use of a supported electrolyte in the
solution, the electrical resistances of
quasi-two-dimensional cells are rather high. Thus the applied potential between
electrodes  affects the electrochemical  process producing a bias for the
impinging particles that arrive at the interface.  It should be
stressed that for the working conditions employed in this work, 
the electrochemical reaction is under mass transfer
control, i.e., the convective-diffusion flux plays an important role 
in determining the morphology of the obtained patterns\cite{g2,g3}.

Figure 1 shows silver patterns, with their final structures of dead trees, 
obtained at two constant cathode-anode potentials.
These patterns consist of an early thin, rough deposit, followed by a 
columnar formation with a preferred growth along the direction of 
the electric field. After a certain deposition time, which decreases 
as  $\Delta E _{c-a}$  is shifted
negatively, separated tree-like formations resulting in a dense
branching with the same preferred growth direction, partially
screening the growth of neighbor columns, can be
observed.  The
morphological characteristics of both aggregates are
significantly different. Thus, the density of trees and the 
aggregate apparent density seem to increase as 
$\Delta E _ {c-a} $ is changed from $-0.80$V to -$1.20$V.

The analysis of the internal structure of the growth patterns is performed
following the procedure outlined in reference \cite{trees}.
Pointing our attention to dead
trees of size \( s \) (\(s\) is the number of particles belonging to the tree),
one has that both, the rms height (\( h_{s} \)) and the rms width (\( w_{s} \))
of trees obey simple power-laws given by  

\begin{eqnarray}
h_{s}\propto s^{\nu _{\parallel }} ,  \label{Eq.1}  
\end{eqnarray}  

\noindent and
\begin{eqnarray}
w_{s}\propto s^{\nu _{\perp }},  \label{Eq.2}
\end{eqnarray}  

\noindent where \( \nu _{\parallel } \) and \( \nu _{\perp } \) are the
correlation length exponents parallel and perpendicular to the main growing
direction of the aggregate \cite{MM}, respectively.

Furthermore, during the competition between trees along the
evolution of the aggregate, the existence of
large neighboring trees inhibit the growing
of smaller ones. This competing process ultimately leads to the death of some
trees that become frozen within the underlying aggregate. These prevailing
large trees continue the competition within more distant trees in a dynamic
process. Since this situation takes place on all scales, it is reasonable to
expect that the tree size distribution ($n_s$) should also exhibit a 
power-law behavior, so that

\begin{eqnarray}
n_{s}\sim s^{-\tau },  \label{Eq.3}
\end{eqnarray}    
\noindent where \( \tau  \) is an exponent.

The evaluation of $h_{s}$, $w_{s}$ and $n_{s}$ was performed using  ten different
samples for each set of experimental conditions. Thus,    
these quantities were evaluated using 1056 trees for $\Delta E_{c-a} = - 0.80 $ V and
1260 trees for $\Delta E _ {c-a} = -1.20 $ V.   
In these measurements, the tree size $s$ was taken as the number of pixels
belonging to the tree. 
The dependences of $h_{s}$ and $w_{s}$ on $s$ are shown in the log-log
plots of figures 2 and 3, respectively. 
In these figures, each point has been obtained averaging over the number of trees 
with size $s$  inside a range $s^{*}-ds,s^{*}+ds$. Thus, the quoted error bars 
only account fot the method applied to that procedure, and the total error is expected to be 
greater.

 It is worth mentioning that,
as follows from these figures, in most cases 
power-laws extends over more than two decades. Also, plots for both
values of $\Delta E_{c-a}$  are almost parallel, indicating the same 
underlying scaling behavior. For $\Delta E_{c-a} = - 0.80 $ V, the 
best fit of the data gives $\nu _{\parallel } = 0.63 \pm 0.03$ and
$\nu _{\perp }=0.46 \pm 0.06$, while for $\Delta E _ {c-a} = -1.20 $ V
one has $\nu _{\parallel } = 0.62 \pm 0.03$ and $\nu _{\perp }=0.46 
\pm 0.06$. Notice, that the 
error in the estimation of the exponents is the standard one, that 
follows from the fitting data.

\begin{figure}
\includegraphics[width=8cm,height=7cm,angle=0]{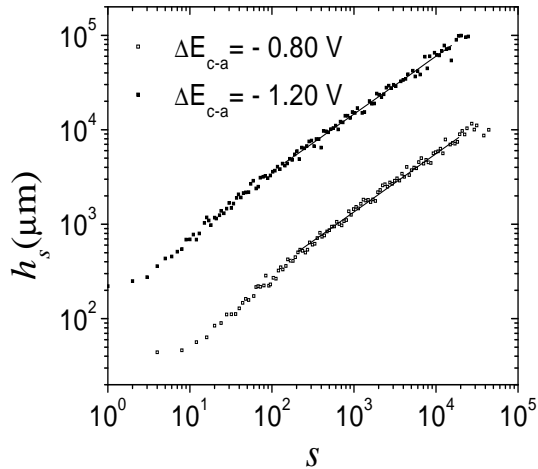}
\caption{Log-log plots of rms height of the trees as a function of
the tree size \( s \). 
The rms height is measured in $\mu m$ and the size $s$ is given by the number 
of pixels of each tree. The result at $\Delta E _ {c-a} = -1.20 $ V has been
shifted by a factor 10 for clarity.}
\label{Figure2}
\end{figure}

\begin{figure}
\includegraphics[width=8cm,height=7cm,angle=0]{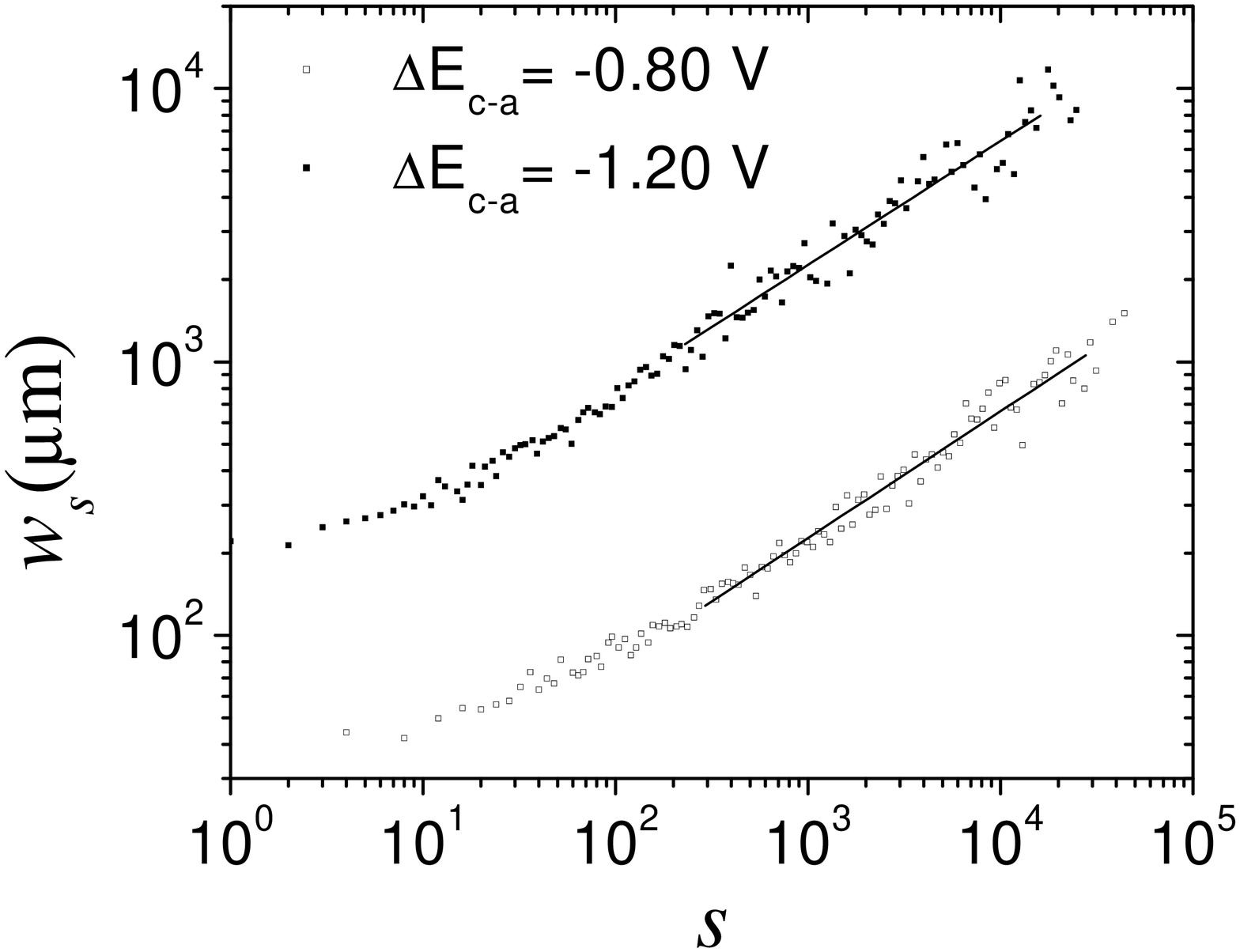}
\caption{Log-log plots of rms width of the trees as a function of
the tree size \( s \). 
The rms width is measured in $\mu m$ and the size $s$ is given by the number 
of pixels of each tree. The result at $\Delta E _ {c-a} = -1.20 $ V has been
shifted by a factor  10 for clarity.}
\label{Figure3}
\end{figure}

On the other hand, the tree size distributions measured for both values
of $\Delta E _{c-a}$ (see figure 4) are fully consistent with
equation (3). Log-log plots of $n_s$ versus $s$ indicate again 
the same scaling behavior, irrespective of $\Delta E _{c-a}$. 
The best fit of the data gives $ \tau  =1.37 \pm 0.04 $. As in the case of $\nu _{\parallel }$ and
 $\nu _{\perp }$ the error only reflect the statistical error.

For self-affine aggregates
with geometrical dimension D growing on a \(d \)-dimensional 
substrate, it has been already established that \cite{MM}

\begin{eqnarray}
\tau =2-\nu _{\parallel }(D-d).  \label{Eq.4} 
\end{eqnarray} 

\begin{figure}
\includegraphics[width=8cm,height=7cm,angle=0]{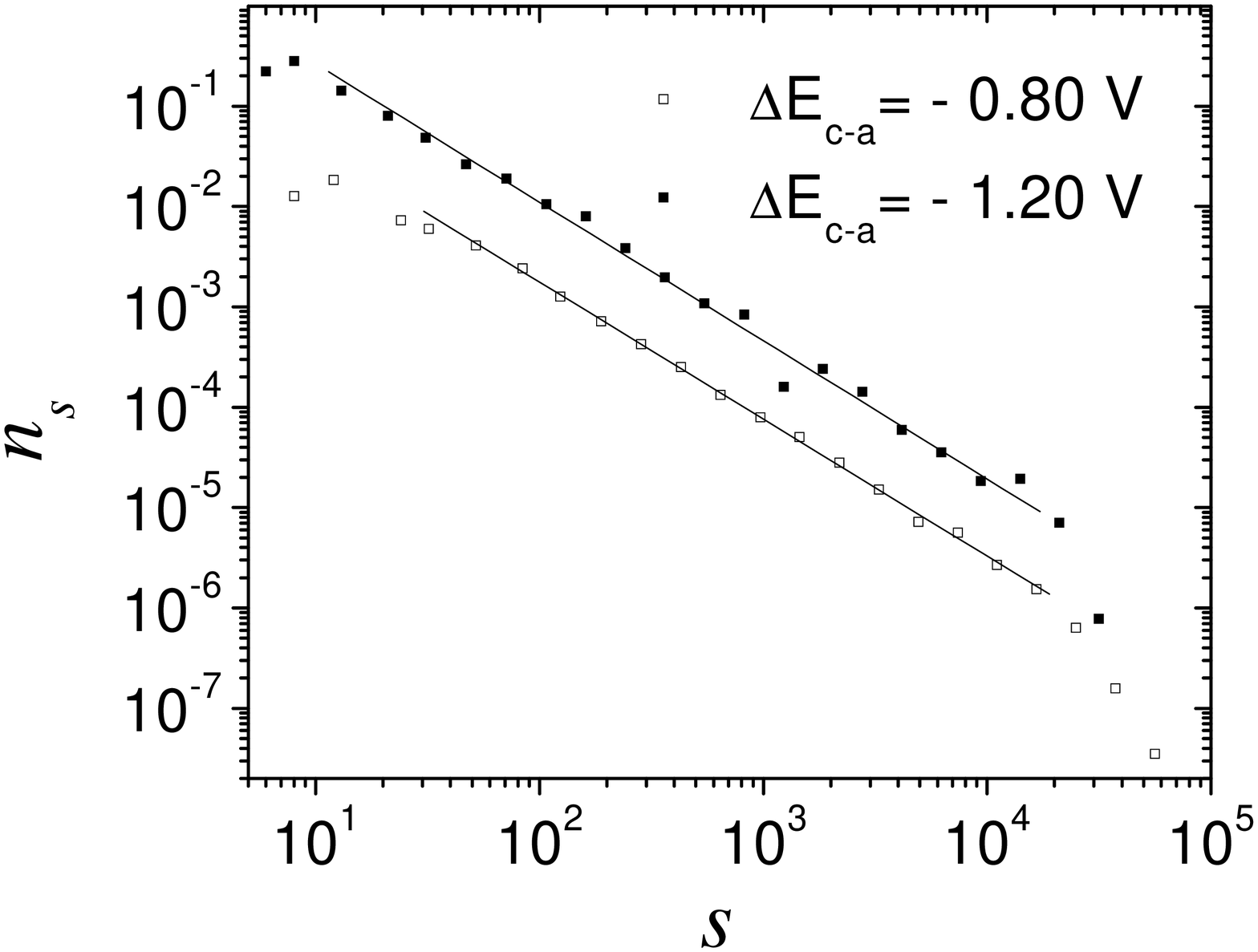}
\caption{Log-log plots of the tree size
distribution, where the size $s$ is given by the number of pixels of 
each tree. The result at $\Delta E _ {c-a} = -1.20 $ V has been shifted by a
factor 10 for clarity.}
\label{Figure4}
\end{figure}

For the experimental conditions used in this work one has $D=2$ and
$d=1$,  therefore equation (4) becomes

\begin{eqnarray}
\tau +\nu _{\parallel }-2=0.  \label{Eq.5} 
\end{eqnarray} 

In principle, one lacks any physical argument to expect that a 
compact (constant apparent density) aggregate may result from the addition of
compact trees.  In fact, the volume of trees of size \( s \) (\( v_{s}
\)) scales, for any dimension, as

\begin{eqnarray}
 v_{s}\sim h_{s}w_{s}^d\sim s^{\nu _{\parallel }+d\nu _{\perp }}, 
\label{Eq.6}
\end{eqnarray}

\noindent and therefore, it is useful to define a relationship $v_s \sim
s^{\pi}$ between the volume and the number of particles, 
where \( \pi  \) is an exponent, so that

\begin{eqnarray}
\nu _{\parallel }+d\nu _{\perp }=\pi, 
\label{Eq.7}
\end{eqnarray} 

\noindent where for compact trees one has \( \pi =1 \) while for
non-compact (fractal) trees  $\pi > 1$.

Finally, identifying the  correlation length perpendicular
 to the main growing direction
  with \( w_{s}
\) and the time with \( h_{s} \), one has that the dynamic exponent
$z = \alpha/\beta$ is given by \cite{Krug} 

\begin{eqnarray}
z=\nu _{\parallel }/\nu _{\perp }. 
\label{Eq.8}
\end{eqnarray} 

Depending on the operating conditions, metal electrodeposition processes involve a 
number of competitive effects that are considered in both non-local and local models used 
to describe growth pattern kinetics. Thus, when the working conditions involve both 
electromigration and activated electron transfer limited kinetics, as it appears to be the case 
of zinc electrodeposition, the value $\tau \simeq 1.54$ has been obtained \cite{zinc}. This figure has been 
interpreted by the diffusion-limited aggregation (DLA) model \cite{DLA}. 
For our working conditions, where silver electrodeposition takes place under almost 
exclusively convective-diffusion rate control (negligible electromigration and electron 
transfer overvoltage), the exponents resulting from the experimental data are assembled in 
Table 1. 
It should be stressed that the formulation of a detailed model describing the complex processes 
involved upon silver electrodeposition is far beyond the aim of this report. 
However, well established concepts in the field of statistical physics, that are very 
useful in the study of phase transition, self-similar and self-affine structures as 
in the present case, allow us to rationalize even very complex systems into few 
universality class. Of course, systems within the same universality share the same 
exponents. For this reason, it is useful to check experimentally evaluated exponents 
to those of the simplest models representing each universality.
Thus, for the sake of comparison, the values of the exponents predicted by the Edward 
Wilkinson (EW) and the Kardar-Parisi-Zhang (KPZ) models are included in Table I. 
The values of these exponents do not depend on $\Delta E_{c-a}$.
 Considering error bars, values of $\nu _{\perp }$ and $z$ are very 
close to those expected from the KPZ universality class. It is worth noting that from our 
experiments the value $\tau = 1.37$ (4) approaches that of the KPZ model instead of $\tau \simeq 1.55$, 
predicted by the DLA model. Besides, the value of $\pi \simeq 1$ indicates that trees behave as 
constant apparent density, compact objects (non fractal). To check the consistency of the 
method employed, the relationship given by equation (5) was evaluated using the 
experimental exponents obtaining a good agreement with the theory.

Summing up, the application of the internal structure analysis 
based method to experimental data allows us to conclude that the
evaluation of treeing statistics is a suitable method
for the characterization of growing aggregates. Furthermore, the exponents
derived from electrodeposited silver patterns are consistent with 
those corresponding to the KPZ universality class. It is expected 
that the statistical analysis of bulk structures may become a
useful tool for the characterization of growing aggregates,
particularly in those cases where the topology of the interface  
is not accessible experimentally.

{\bf Acknowledgments}: This work was supported by CONICET, 
UNLP and ANPCyT (Argentina).

\vspace{-0.25cm}
\begin{table}
\caption {Scaling exponents $\nu _{\perp }$, $\nu _{\parallel}$, 
$\tau$ determined from electrodeposited silver patterns and the  
predictions corresponding to both the KPZ and EW universality
classes are shown in the \( 2^{nd} \), \( 3^{rd} \) and 
\( 4^{th} \) columns, respectively. A test of equation (5) is shown
in the \( 5^{th} \) column. Values of \( \pi  \) (equation (7)) and 
$z$ (equation (8))  are shown in the \( 6^{th}\) and \( 7^{th} \)
columns, respectively.\\  } 
\end{table}

\vspace{0.3cm}
{\centering \begin{tabular}{|c|c|c|c|c|c|c|}
\hline 
System &
\(\nu _{\perp }\)&
\( \nu _{\parallel }\)&
\( \tau\)&
\( \tau +\nu _{\parallel }-2\) &
\( \pi \)&
\( z \)    \\
\hline 
\hline 
Experimental ($\Delta E _ {c-a} = -0.80 $ V)&
0.46(6)&
0.63(3)&
1.37(4)&
0.00(5)&
1.09(7)&
1.4(2)   \\
\hline 
\hline 
Experimental ($\Delta E _ {c-a} = -1.20 $ V)&
0.46(6)&
0.62(3)&
1.37(4)&
0.01(5)&
1.08(7)&
1.4(2)  \\
\hline 
\hline 
KPZ universality class&
0.40&
0.60&
1.40&
0&
1&
3/2  \\
\hline 
\hline 
EW universality class&
0.35&
0.65&
1.35&
0&
1&
2  \\
\hline 
\end{tabular}\par}
\vspace{0.3cm}

\end{document}